 \documentclass[final,3p,11pt,times,floatfix]{elsarticle}
\usepackage{amssymb}
\usepackage{alltt}

\journal{Comp.~Phys.~Comm.}

 \usepackage{hyperref}
 \hypersetup{
    bookmarks=true,         
    unicode=false,          
    pdftoolbar=true,        
    pdfmenubar=true,        
    pdffitwindow=false,     
    pdfstartview={FitH},    
    pdftitle={SIMLA_CompPhysCommun},   
    pdfauthor={D. G. Green},     
    pdfsubject={SIMLACompPhysComm},   
    pdfcreator={D.~G.~Green and C.~N.~Harvey},   
    pdfproducer={Producer}, 
    pdfkeywords={keywords}, 
    pdfnewwindow=true,      
    colorlinks=true,       
    linkcolor=blue,          
    citecolor=blue,        
    filecolor=magenta,      
    urlcolor=blue           
}

\begin{document}

\begin{frontmatter}

\title{SIMLA: Simulating laser-particle interactions via classical and quantum electrodynamics\tnoteref{tnote1}}
\tnotetext[tnote1]{The associated computer program and corresponding manual will be made available on the CPC library}

\author{D.~G.~Green \fnref{dg_footnote}}
\ead{dermot.green@balliol.oxon.org}
\author{C.~N.~Harvey\corref{cor1}\fnref{chris_footnote}}
\address{Centre for Plasma Physics, School of Mathematics and Physics, Queen's University Belfast, Belfast, Northern Ireland, BT71NN, United Kingdom.}
 
\fntext[dg_footnote]{Present address: Joint Quantum Centre (JQC), Durham/Newcastle, Department of Chemistry, Durham University, South Road, Durham, DH1\,3LE.}

\fntext[chris_footnote]{Present address: Department of Applied Physics, Chalmers University of Technology, SE-41296 Gothenburg, Sweden.}
\ead{cnharvey@physics.org}
\cortext[cor1]{Corresponding author. Tel: +46 31-772 10 00}

\begin{abstract}
We present the Fortran code SIMLA, which is designed for the study of charged particle dynamics in laser and other background fields.
This can be done classically via the Landau-Lifshitz equation, or alternatively, via the simulation of photon emission events determined by strong-field quantum-electrodynamics amplitudes and implemented using Monte-Carlo type routines.
Multiple laser fields can be included in the simulation and the propagation direction, beam shape (plane wave, focussed paraxial, constant crossed, or constant magnetic), and time envelope of each can be independently specified.

\end{abstract}

\begin{keyword}
intense laser-electron interactions, strong-field quantum electrodynamics, radiation reaction.
\PACS 41.60.-m \sep 12.20.-m \sep 03.50.De \sep 52.38.Ph \sep 41.75.Jv 
\end{keyword}

\end{frontmatter}

\section{Introduction}\label{sec:intro}
Recent technological advances have led to the development of lasers of unprecedented powers and intensities.  The current record of $2\times 10^{22}$\,Wcm$^{-2}$ was set by the HERCULES laser in 2008~\cite{Yanovsky:2008}.  It is expected that this will be routinely surpassed when a number of new facilities come online, such as the European `Extreme Light Infrastructure' (ELI)~\cite{ELI} and the Russian ExaWatt Centre for Extreme Light Studies (XCELS)~\cite{XCELS}.  These facilities aim to reach intensities of the range $10^{23}-10^{25}$\,Wcm$^{-2}$, allowing the probing of fundamental physics in the ultra-high intensity regime.  In particular, these lasers will enable an investigation of intensity effects in strong field quantum electrodynamics (QED) (see, e.g., \cite{DiPiazza:2011tq,Heinzl:2008an}).

These new facilities will, in part, achieve such intensities by strongly focussing their beams.  As a result the particle dynamics, and resulting emission spectra, will be very difficult to determine analytically.  Thus one is forced to resort to numerical schemes in order to simulate the interactions.  Recently there have been a number of publications presenting modifications to particle-in-cell (PIC) codes to include QED processes (see, e.g., \cite{Bell:2008zzb,Sokolov:2010am, Elkina:2010up, Ridgers2014273}).  While having the particle-particle capability of a PIC based codes is certainly important for many applications, there are also many situations where the mutual interactions between particles in the field is not of interest and can be neglected.  In the case of experiment, a typical example might be a beam of high-energy particles colliding with a laser pulse.  Moreover, the dynamics of single isolated particles are of theoretical interest, both to better understand the fundamental physics  of particles in these background fields (see for example \cite{PhysRevA.85.013412,PhysRevLett.113.014801,PhysRevLett.112.164801}) and in order to test and refine the numerical models used (see \cite{numericalspectrapaper}).  Although PIC schemes can still be used in such cases, we believe it is beneficial to have a dedicated single-particle code that can run on an ordinary desktop computer, and used by a wider audience without reliance on the specialist knowledge and super-computing infrastructure required to run a PIC code. In addition, such a dedicated single-particle code can also be made more efficient than its PIC counterpart, since it is possible to dynamically change the grid spacing during a run.  

In this paper we present our single-particle code SIMLA. It is a modular Fortran code that can be run on a modest desktop computer to calculate the dynamics of a charge in laser fields. SIMLA accounts for the effects of radiation reaction via either numerical solution of e.g., the Landau-Lifshitz equation (for classical particles), or via Monte-Carlo simulation using QED emission amplitudes (for quantum particles). Multiple laser fields can be included in the simulation and the propagation direction, beam shape (plane wave, focussed paraxial, constant crossed, or constant magnetic), and time envelope of each can be independently specified \footnote{In the current version up to nine independent laser fields can be specified. The code could easily be adapted to handle a larger number, if required.}.

The structure of the paper is as follows. In Section \ref{sec:theory} we present a brief overview of the theory of charged particles in laser fields, and its specific implementation in the SIMLA code. We consider the case of the motion of non-radiating classical particles, radiating classical particles, and the stochastic effects of photon emission from quantum particles. In Section \ref{sec:testproblems} a number of test problems are outlined that provide a tutorial introduction to the basic functionality of the code. We conclude with a summary in Section \ref{sec:conclusion}.\\\\

\section{Theory of charged particles in laser fields and its implementation in SIMLA: radiation reaction and stochastic dynamics}\label{sec:theory}

\subsection{Classical theory}
\subsubsection{Non-radiating classical motion}
In the classical case where radiation damping effects are neglected, the motion of a charged particle in an electromagnetic background field is described by the Lorentz-force equation,
\begin{eqnarray}
\frac{d\mathbf{p}}{dt} &=&q\left(\mathbf{E}+\mathbf{v}\times \mathbf{B}\right), \label{LF}
\end{eqnarray}
where $\mathbf{p}=\gamma m\mathbf{v}$ is the relativistic momentum expressed in terms of the $\gamma$-factor, velocity $\mathbf{v}$, charge of the particle $q$ and the particle rest mass $m$ (and we are working in units such that $\hbar=c=1$).  The electric and magnetic field components $\mathbf{E}$ and $\mathbf{B}$ can be arbitrary functions of space and time.
In this non-radiating classical case, the SIMLA code propagates the particle throughout the field by solving Eq.~(\ref{LF}) to determine the acceleration.  Although (\ref{LF}) is only expressed in terms of three-vectors, the code maintains covariance by enforcing the mass-shell condition $p^2=m^2$ when calculating $\gamma$ (for an alternative solution to this problem we refer the reader to \cite{PhysRevD.83.076013}).

In the case of a time-dependent field, such as a laser pulse, it is conventional to specify the field strength using a dimensionless parameter defined with reference to a probe particle.  This is $a_0\equiv qE/\omega m $, where $\omega$ is the frequency of the background field and $E$ is the {\it maximum} strength of the electric field.  

\subsubsection{Radiation reaction}
Both classical and quantum electrodynamics dictate that a particle that is accelerated by a background field will radiate.  If the acceleration is strong, this radiation can lead to a significant reduction in the energy and momentum of the particle.  
In the classical case, the spectrum of radiation is continuous (the quantum case is discussed below), 
and the effect of this `radiation reaction' on the particle dynamics can be included by means of a correctional term to the Lorentz Force equation (Eq.~(\ref{LF})). 
Determining the `exact' form of this correction is, however, non-trivial. It has been the subject of study for over 100 years, and remains one of the most fundamental problems in electrodynamics.
The most common starting point to describe the radiation reaction is the Lorentz-Abraham-Dirac equation, which is obtained by solving the coupled Lorentz and Maxwell's equations \cite{Lorentz:1905,Abraham:1905,Dirac:1938nz}.  
However, this equation is notorious due to defects such as pre-acceleration and (unphysical) runaway solutions.  One of the most common ways of circumventing these problems is to adopt the perturbative approximation of Landau and Lifshitz \cite{LLII}, so that the equation of motion becomes
\begin{eqnarray}
\frac{d\mathbf{p}}{dt}=\mathbf{f}_\textrm{L}+\mathbf{f}_\textrm{R},
\end{eqnarray}
where $\mathbf{f}_\textrm{L}$ is the Lorentz force (\ref{LF}) and the radiative correction term 
\begin{eqnarray}
\mathbf{f}_\textrm{R}&=&-\left( \frac{2}{3} \frac{q^3}{4\pi m}  \right)\gamma\left[ \left( \frac{\partial}{\partial t}+\mathbf{v}\cdot\nabla\right)\mathbf{E}+\mathbf{v}\times \left( \frac{\partial}{\partial t}+\mathbf{v}\cdot\nabla\right)\mathbf{B}\right]\nonumber\\
&&+\left( \frac{2}{3} \frac{q^4}{4\pi m^2}  \right) [(\mathbf{E}+\mathbf{v}\times\mathbf{B})\times\mathbf{B}+(\mathbf{v}\cdot\mathbf{E})\mathbf{E}]\nonumber\\
&&-\left( \frac{2}{3} \frac{q^4}{4\pi m^2} \right) \gamma^2 [(\mathbf{E}+\mathbf{v}\times\mathbf{B})^2-(\mathbf{v}\cdot\mathbf{E})^2]\mathbf{v}.\label{LL}
\end{eqnarray}
Equation (\ref{LL}) is valid when the radiative reaction force is much less than the Lorentz force in the instantaneous rest frame of the particle.  We note that there are numerous alternative equations in the literature (see, for example, \cite{Sokolov:2009,O'Connell:2012ee}) and it is still an open problem as to which is the correct formulation.  However, the Landau-Lifshitz equation has, along with some others, recently been shown to be consistent with quantum electrodynamics to the order of the fine-structure constant $\alpha$ \cite{0038-5670-34-3-A04,Ilderton:2013tb}
\footnote{Alternative equations of motion may be implemented simply by replacing or modifying the particle pusher module.}.

It should be noted that the first term (derivative term) of Eq.~(\ref{LL}) is significantly smaller than the other two since it is only linear in the field strength whereas the other terms are quadratic.  It is found that in almost all cases the contribution from this term is negligible and so, to increase computational speed, it is not evaluated by the code.

\subsubsection{Classical emission spectrum}
The expression describing the classical radiation emission of a particle can be found in most textbooks on electrodynamics (see, e.~g.~\cite{jackson}).  However, in the SIMLA code we adopt the less common covariant formulation which can be found, for example,  in~\cite{Mitter:1998}.  The radiated energy is given by the zero component of the four-momentum of the emitted radiation
\begin{eqnarray}\label{P0}
P^0\equiv \int \rho \omega^\prime d\omega^\prime d\Omega ,  
\end{eqnarray}
where
\begin{eqnarray}
\rho\equiv \frac{d^2N_\gamma}{d\omega^\prime d\Omega}=-\frac{\omega^\prime}{16\pi^3}j(k^\prime)\cdot j^\ast (k^\prime),\label{spectraldensity}
\end{eqnarray}
is the classical spectral density, which gives the `number of photons', $N_\gamma$, radiated per unit frequency per unit solid angle $d\Omega$.
Here 
\begin{eqnarray}
j^{\mu}(k^{\prime})=e \int d\tau ~u^\mu (\tau) e^{-ik^\prime\cdot x(\tau )}
\end{eqnarray}
is the Fourier transform of the electron current and $k^{\prime\mu}=\omega^\prime (1,\sin\theta \cos\phi,\sin\theta\sin\phi,\cos\theta)$ is the vector of the emitted radiation.
However, the boundary conditions can result in an infra-red divergence which means it is necessary to regulate the integral; see~\cite{Dinu:2012} for a fuller discussion.  The resulting expression is 
\begin{eqnarray}
j^\mu(k^\prime)=e\int  e^{-ik^\prime\cdot x(\tau)} \frac{d}{d\tau}\left(\frac{u^\mu}{ik^\prime\cdot u}\right).\label{j_noboundary}
\end{eqnarray}
Performing an integration by parts we find
\begin{eqnarray}
j^\mu(k^\prime) = e\frac{u^\mu}{ik^\prime\cdot u}e^{-ik^\prime\cdot x(\tau)}\bigg|_{\tau=-\infty}^{\tau=+\infty}  +e\int d\tau ~u^\mu(\tau)e^{-ik^\prime\cdot x(\tau)},
\end{eqnarray}
which is (\ref{j_noboundary}) with the addition of some boundary terms.  

The evaluation of the oscillatory integrals in (\ref{P0}) can be a very numerically expensive procedure, especially for large field intensities.
However, the nonlinear time grid improves efficiency of the integration. (Although, at the same time, this makes it more technically difficult to efficiently implement higher-order methods.  SIMLA currently uses a trapezoidal method to evaluate the integrals.)

\subsection{Strong-field quantum electrodynamics}
As well as classical calculations, the code can calculate the trajectories and resulting emission spectra of the particles using a Monte Carlo QED method, which we will now discuss.

We begin by introducing the dimensionless and invariant  `quantum efficiency' parameter for the particle $\chi_e\equiv \sqrt{(F^{\mu}_{\phantom{\mu}\nu}p^\nu)^2}/m^2\sim\gamma E/E_{\rm cr}$, where $E_{\rm cr}=1.3\times10^{16}$\,Vcm$^{-1}$ is the QED `critical' field (`Sauter-Schwinger' field)~\cite{QEDcriticalfield1,QEDcriticalfield2,QEDcriticalfield3}.  This can be interpreted as the work done on the particle by the laser field over the distance of a Compton wavelength, and is thus a measure of the importance of quantum effects for a given set of parameters.  
In the limit $a_0\gg 1$ the size of the radiation formation region is of the order $\lambda/a_0\ll\lambda$, where $\lambda=2\pi/\omega$ is the laser wavelength~\cite{Ritus:1985}.   
Thus the laser varies on a scale much larger than the formation region and so can be approximated as locally constant and crossed, allowing us to determine the probability of photon emission using the differential rate~\cite{Ritus:1985}
\begin{eqnarray}\label{eqn:dGam}
{d\Gamma}&=&\frac{\alpha m}{\sqrt{3}\pi\gamma\chi_e}
\left[\left( 1-\eta+\frac{1}{1-\eta}\right) K_{2/3}(\tilde{\chi})\right. 
\left.-\int_{\tilde{\chi}}^\infty dx K_{1/3}(x)\right] {d\chi_\gamma},
\end{eqnarray}
where $K_\nu$ is the modified Bessel function of order $\nu$, 
$\eta\equiv \chi_{\gamma}/\chi_e$, $\tilde{\chi}\equiv 2\eta/\left[3\chi_e\left(1-\eta\right)\right]$, 
and we have introduced the analogous invariant parameter $\chi_\gamma\equiv \sqrt{(F^{\mu}_{\phantom{\mu}\nu}\kappa^\nu)^2}/m^2$ for the emitted photon with momentum $\kappa^\nu$.  

The QED routines operate in the following manner.
The code implements its classical particle pusher to propagate the particle through the field via the Lorentz equation. 
At each time step a uniform random number $r\in[0,1]$ is generated, and emission deemed to occur if the condition $ r\leq \Gamma dt$ is satisfied, under the requirement $\Gamma dt\ll1$. 
Such an event generator has been used in Ref.~\cite{Elkina:2010up}.
Note that during the simulation $d\Gamma$ (and thus $\Gamma$) is a time-dependent quantity owing to the effect of the temporally varying laser pulse and electron motion. 
Given an emission event, the photon $\chi_{\gamma}$ is determined as the root of the sampling equation $\zeta={\Gamma(t)}^{-1} \int_{0}^{\chi_{\gamma}}d\Gamma(t)$, where $\zeta$ is a uniform random number $\zeta\in[0,1]$ (In actual fact, the integral is performed from a lower limit $\varepsilon \sim 10^{-5}$, rather than zero: 
the emission of soft photons of energy below this cut off will not appreciably affect the particle motion (see e.g., Ref.~\cite{Duclous}).)
Next, we calculate the photon momentum from $\chi_{\gamma}$ assuming that the emission is in the direction of motion of the particle.
This is valid for $\gamma\gg 1$, since in reality the emissions will be in a cone of width $\gamma^{-1}$~\cite{jackson, Harvey:2009ry}. 
Finally, the particle momentum is updated and the simulation continues by propagating the particle via the Lorentz equation to the next time step.

This method of implementing the emission process via statistical routines is similar to those used in a number of recently developed PIC codes for the modeling of QED cascades, see e.g.,~\cite{Bell:2008zzb,Sokolov:2010am, Elkina:2010up}.  The validity of using such a method has recently been tested in \cite{numericalspectrapaper} where it was found that it accurately reproduces the correct photon spectra in all but a few special cases.

Note that when $\chi_e\gtrsim1$ quantum effects dominate and pair production can occur. To maintain accessibility and efficiency, this version of the SIMLA code is strictly a single-particle code. It is therefore not designed to run in such a regime, and will abort if the situation occurs \footnote{Note that for multi-particle codes, pair-production events can be implemented via similar Monte-Carlo methods to those described above for the photon emission, see, e.g., \cite{Duclous}.}

\subsection{Further information about the workings of the code}\label{sec:furtherinfo}
\begin{figure*}[ht!]
\centering
\includegraphics*[width=0.75\columnwidth]{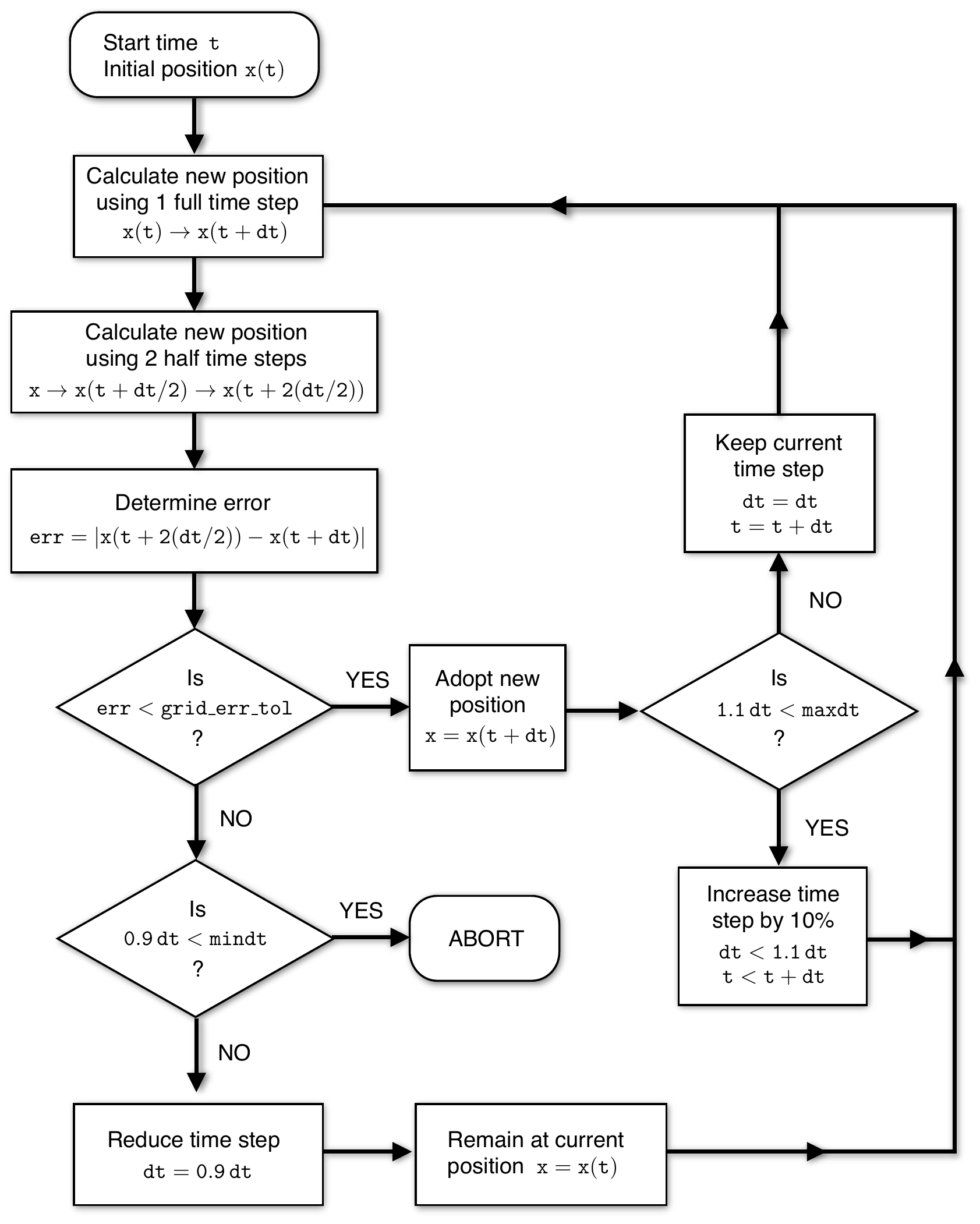}
\caption{Flow chart showing the operation of the adjustable time step.  \label{fig:timestep} }
\end{figure*}

One important feature of the code is that it uses an adjustable time step.  This is one reason why, in cases where collective effects can be neglected, the code is superior to PIC codes, where such a feature is not possible.  The user specifies in the input files the maximum permissible error in the norm of the spatial coordinates that can be introduced at each time step.  The code then propagates the particle through the simulation, making continuous adjustments to the time step such that the error condition is maintained.  Thus the grid points are not uniform; there will be very few time steps in regions where the background field is weak, while there will be many in more intense/complex regions of the field.  We have found that the implementation of the adjustable grid typically decreases the computation time by around two orders of magnitude compared to a fixed size grid.  A flow chart showing how the method is implemented is given in Fig.~\ref{fig:timestep}.

\begin{table}
\caption{The units used by the code.  Note that the input parameters should be in the units specified in Tables \ref{tab:particleinput}, \ref{tab:inputfields}, \ref{tab:inputsetup}.  \label{tab:units}}
\begin{tabular}{ l | l }
  \hline    
  Quantity & Units \\
  \hline                   
  Length & eV$^{-1}$  \\
  Time & eV$^{-1}$  \\
  Mass & eV  \\
   Charge & $e/\sqrt{4\pi\alpha}$  \\
  \hline  
\end{tabular}
\end{table}

Next we make some remarks about the normalisations adopted in the code.  In codes such as these it is common to use dimensionless units, usually by normalising the field amplitudes in terms of $a_0$.  However, such normalisation was found to be impractical in our case, where we allow multiple fields with different frequencies and also constant fields which do not have a time dependence.  Thus we adopt (Lorentz-Heaviside) natural units, where $\hbar=c=1$.  The resulting units of the various quantities in the code are specified in Table \ref{tab:units}.  For the typical parameters involved in a laser-particle simulation these units are found to result in quantities that are roughly of order unity and so wholly suitable as a numerical implementation.  (Note, however, that these units are {\it not} the same as the units of the input variables which should be defined according to the conventions specified in Tables \ref{tab:particleinput}, \ref{tab:inputfields}, \ref{tab:inputsetup}.) 

Finally, we note that a more thorough overview of the code is given in the accompanying manual.  This includes instructions explaining how to compile and execute the code on different systems.

\section{Test problems}\label{sec:testproblems}
\subsection{Trajectory and classical emission spectrum of electron in a plane-wave field}
As a first example, let us use the code to calculate the trajectory and classical emission spectrum of an electron colliding head on with a pulsed plane-wave field.  The code requires three input files: one defining the initial conditions of the particle(s), one defining the background field(s) and one specifying the remaining parameters that describe, e.g., the simulation volume.

\begin{table}
\caption{The required variables for each particle in the file {`particle\_input.csv'}. \label{tab:particleinput}}
\begin{tabular}{ l  l  l }                    
Pos. & Quantity & Comments \\
\hline
1 & particle label & for user reference only \\
2 & species & e, p, H+ \\
3 & $\theta_0$ (deg) & $\theta_0=\arctan(y_0/x_0)$ \\
4 & $\phi_0$ (deg) & $\phi_0=\arccos(z_0/dist_0)$ \\
5 & $dist_0$ (m) & initial distance from ($x_0,y_0,z_0$)  \\
6 & $x_0$ (m) & target $x$-coord. \\
7 & $y_0$ (m) & target $y$-coord. \\
8 & $z_0$ (m) & target $z$-coord. \\
9 & $\gamma_0$ & initial $\gamma$-factor \\
10 & eq.~of motion & lf, ll, qed \\
11 & output flag & t, x, ct\\
\hline  
\end{tabular}
\end{table}

\begin{figure}[htp!]
\centering
\includegraphics*[width=0.7\columnwidth,clip=true,viewport=30 190 545 595]{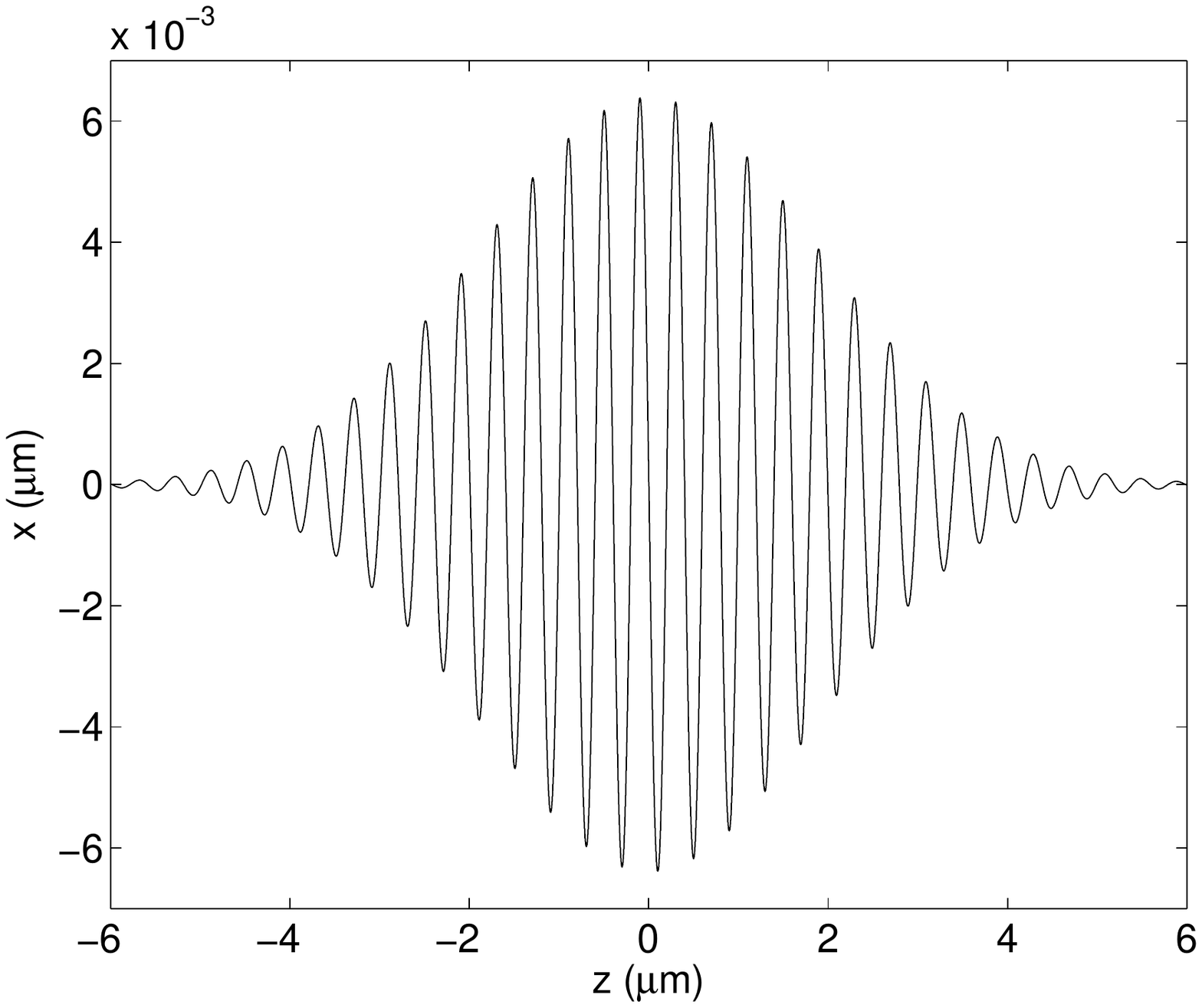}\\
\includegraphics*[width=0.7\columnwidth,clip=true,viewport=30 190 545 595]{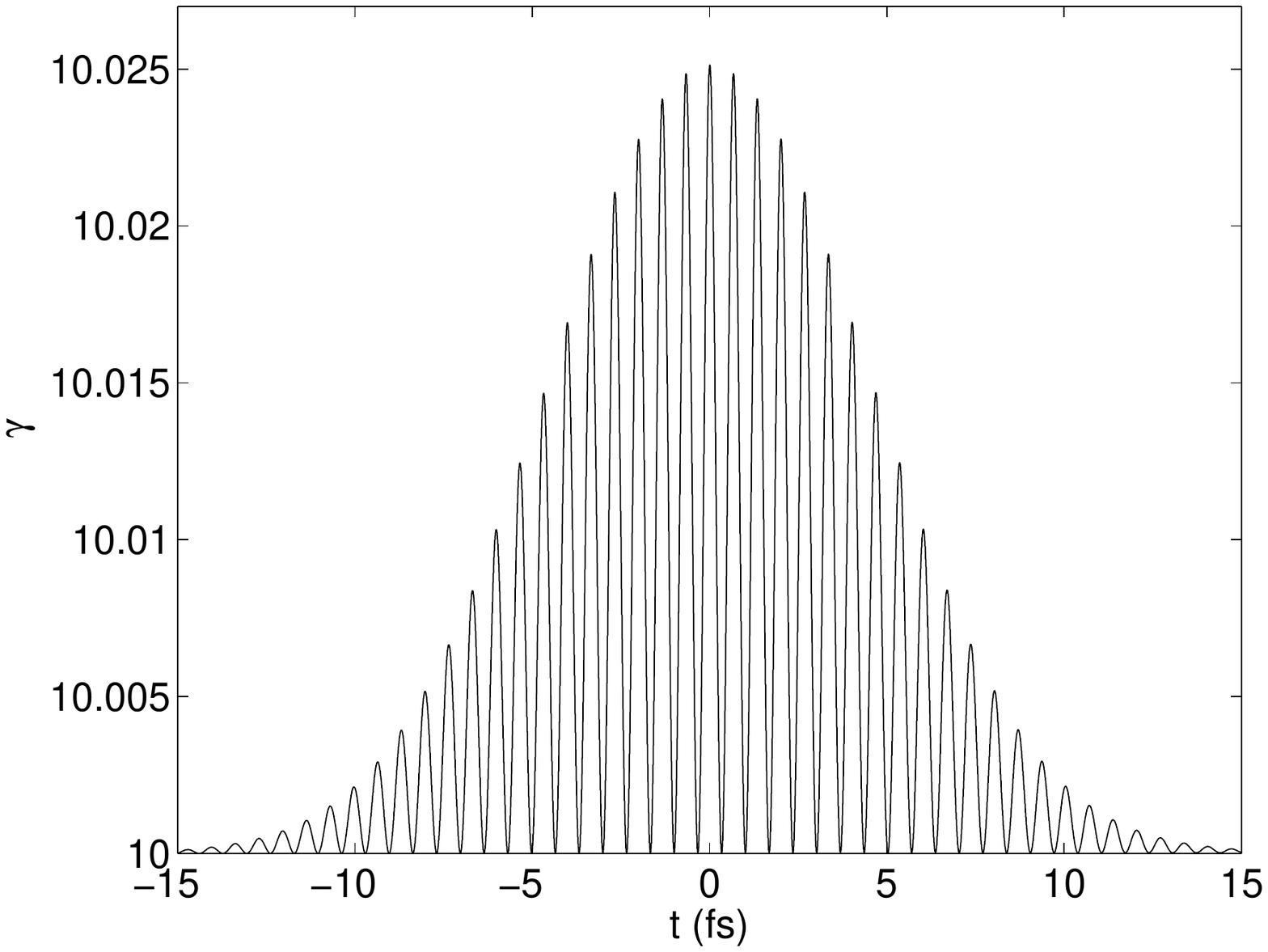}
\caption{The trajectory (top panel) and $\gamma$-factor (bottom panel) for an electron in a head-on collision with a 30\,fs linearly polarised wave of intensity $a_0=1$ and wavelength $\lambda=0.8\,\mu$m. \label{fig:traj} }
\end{figure}
\begin{figure}[htp!]
\centering
\includegraphics*[width=0.7\columnwidth,clip=true,viewport=40 190 550 590]{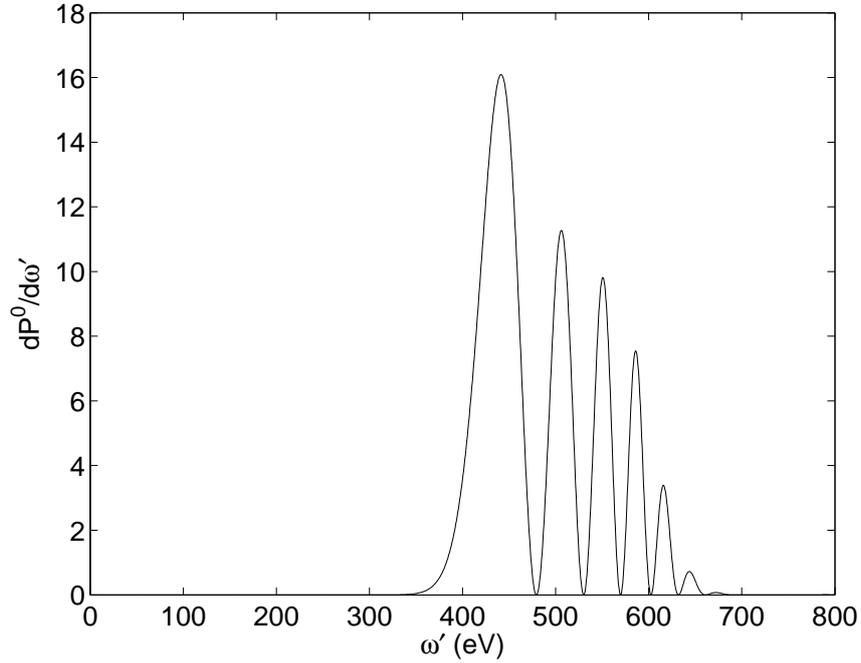}
\caption{The classical emission spectrum calculated using the particle orbit from Fig.~\ref{fig:traj}. \label{fig:spectrum} }
\end{figure}

\begin{figure}[htp!]
\centering
\includegraphics[width=0.7\columnwidth,clip=true,viewport=40 190 540 590]{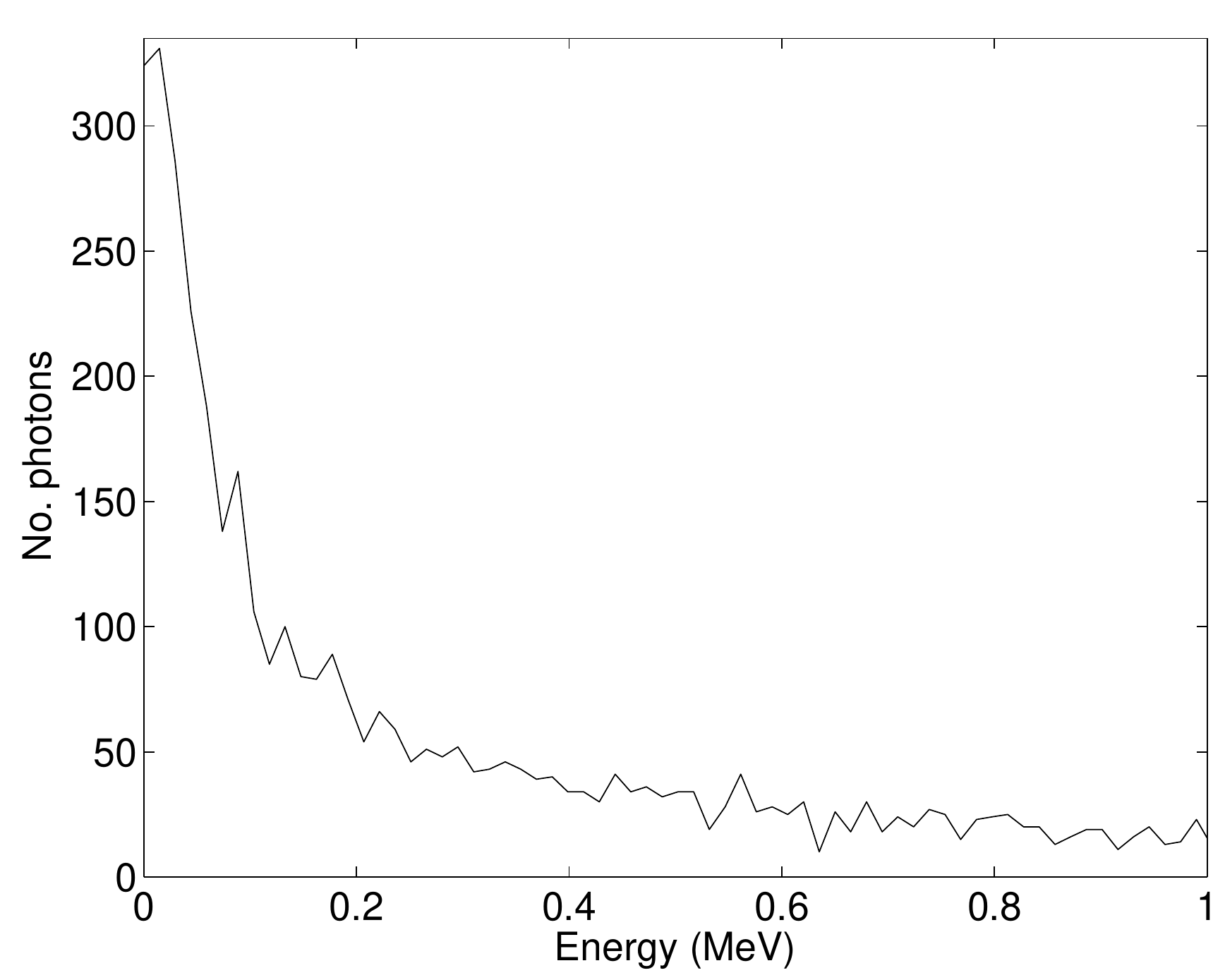}\\[2ex]
\includegraphics[width=0.7\columnwidth,clip=true,viewport=40 190 540 590]{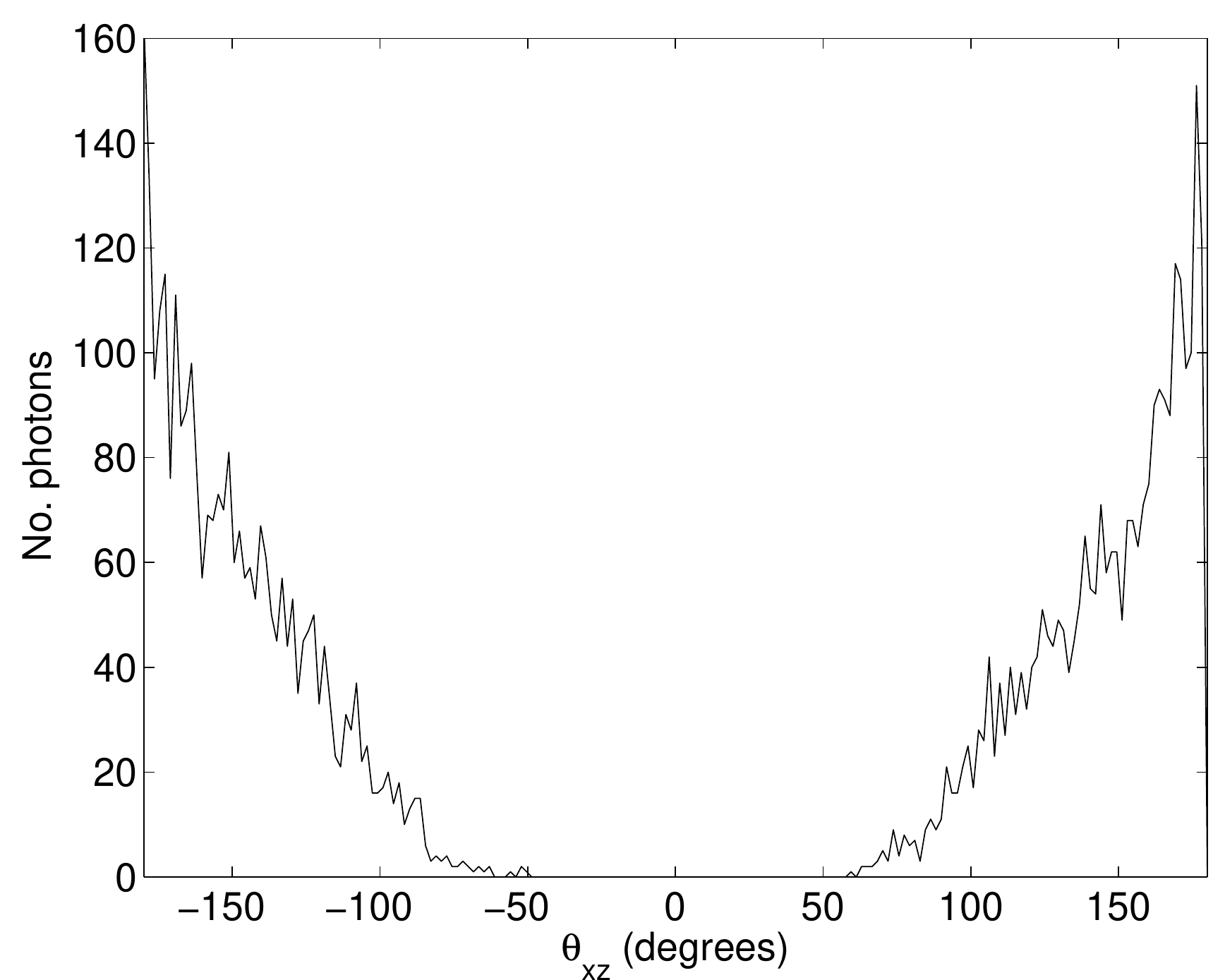}
\caption{The QED photon emission spectrum as a function of photon frequency (top panel), and as a function of emission angle (bottom panel), for an electron of $\gamma_0$=200 in collision with a 30\,fs linearly polarised wave of intensity $a_0=150$ and wavelength $\lambda=0.8\,\mu$m.\label{fig:photon_spectra} }
\end{figure}

We begin by stating the initial conditions of the particle in the file {`particle\_input.csv'}.  This file is different to the other two in that its ordering is important.  The first line of the file is for comments.  The second line can either read {\tt repeatfirstline=on} or {\tt repeatfirstline=off} depending on whether the user wishes to run simulations using each line of particle data in turn or instead wishes to run multiple simulations repeatedly using the data defined in the first line.  The third line must contain an integer specifying the number of runs.  Each subsequent line of the file specifies the initial conditions for each particle in turn.  These lines must consist of 11 entries separated by commas and as defined in Table \ref{tab:particleinput}.  In this case we want an electron initially propagating in the negative $z$-direction so we set $\theta_0=0^\circ$, $\phi_0=180^\circ$.  Specifically, we want it to begin 100\,$\mu$m from the origin and to be initially travelling in a direction aiming for the coordinate origin.  Therefore, we set $dist_0=100 \times 10^{-6}$ and $x_0=y_0=z_0=0$.  The resulting {`particle\_input.csv'} file should take the form
\begin{alltt}
! ...comments 
repeatfirstline=off
1
1,e,0,180,100e-6,0,0,0,10,lf,t
\end{alltt}
where we have set the initial $\gamma_0=10$.  The flag {\tt lf} means that we want to find the trajectory using the Lorentz force equation (\ref{LF}) and the flag {\tt t} tells the code to output the trajectory data to file.

Next we define the background field via the file {`input\_fields.txt'}.  This file is more flexible than the `particle\_input.csv' file and the declarations can appear in any order.  We only want one field -- a linearly polarised plane wave with a 30\,fs Gaussian time envelope -- so we set {\tt no\_fields=1}, {\tt field1=linpw}, {\tt profile1=gauss}, {\tt duration1=30e-15}.  We set the dimensionless intensity to one, {\tt a0\_1=1}, and the wavelength to 0.8\,$\mu$m, {\tt lambda=0.8e-6}.  Finally, to have the field propagating in the positive $z$-direction we set the angles {\tt anglexz1=0}, {\tt angleyz1=0}, {\tt anglexy1=0}.  The full range of options for the {`input\_fields.txt'} file are given in Table \ref{tab:inputfields}.

The file `input\_setup.txt' defines the workings of the code.  
The first thing to understand is that there are two volumes, or \emph{boxes}, associated with a simulation -- the \emph{simulation box} and the \emph{write box}.  The simulation box defines the region of spacetime in which the simulation takes place.  Once the particle leaves the confines of this box the code will terminate.  The write box is a subspace of the simulation box.  When the particle is within the confines of the write box the trajectory and momentum data will be written to file (assuming the output flag for the particle is set to {\tt t} in the `particle\_input.csv' file).  It is possible for the particle to enter and leave the write box multiple times during the simulation; the particle data will only be written to file during those times in which it is inside the box.  In the file `input\_setup.txt' we have to specify the dimensions of both of these boxes.
To define the simulation box as a cube of length 400\,$\mu$m centred around the origin and of maximum time 3000\,fs we add the following lines to the `input\_setup.txt' file
\begin{alltt}
tmax=3000e-15
xmin=-200e-6
xmax=200e-6
ymin=-200e-6
ymax=200e-6
zmin=-200e-6
zmax=200e-6
\end{alltt}
Similarly, we can define the write box to be a 60\,$\mu$m sized subspace of the simulation box, containing the time interval -30\,fs to 30\,fs
\begin{alltt}
tminw=-30e-15
tmaxw=30e-15
xminw=-30e-6
xmaxw=30e-6
yminw=-30e-6
ymaxw=30e-6
zminw=-30e-6
zmaxw=30e-6
\end{alltt}
The code operates with an adjustable time grid, as outlined in Sec.~\ref{sec:furtherinfo}, and use of this feature required specification of additional parameters.  These include the initial, minimum and maximum allowed sizes for the time step (in s) as well as the maximum allowed error over each time interval (defined in terms of the norm of the position vector).  Additionally, to reduce the size of the output files we specify after how many time steps the data should be written to file.  For our problem the entries should take the form
\begin{alltt}
mindt=1e-35
maxdt=0.1e-15
initialdt=1e-20
grid_err_tol=1e-10
writeevery=50
\end{alltt}
Here, the {\tt maxdt} and the {\tt writeevery} are much smaller than we would typically require to make plots of the trajectory only.  This is because we are intending to calculate the classical emission spectrum and, even for the low intensity field we are using here, the evaluation of the integrals in (\ref{spectraldensity}) requires a high degree of precision.  Finally, we choose to use a leapfrog numerical solver by setting {\tt solver=leapfrog}, we specify the output to be in ascii text format by setting {\tt fileformat=txt}, and we choose not to have the background field data outputted by setting {\tt outputintensity=off}.  Note that the full range of options for the `input\_setup.txt' file are given in Table \ref{tab:inputsetup}.

Once the code has been executed the particle trajectory data will be written to the file {`trajectory00001.dat'}.  The Matlab script {`simlaplot.m'} can be executed to plot a selection of the data,  which should correspond to that shown in Fig.~\ref{fig:traj}.  Now, with the trajectory data still loaded into the Matlab memory, we can calculate the classical emission spectrum by running the script {`spectrum.m'}.  The output from this script is plotted in Fig.~\ref{fig:spectrum}.

\subsection{Generating a QED photon spectrum and comparing the orbits of some of the quantum particles with a classical radiating particle}
\begin{figure}[htp!]
\centering
\includegraphics[width=0.7\columnwidth,clip=true,viewport=40 190 545 590]{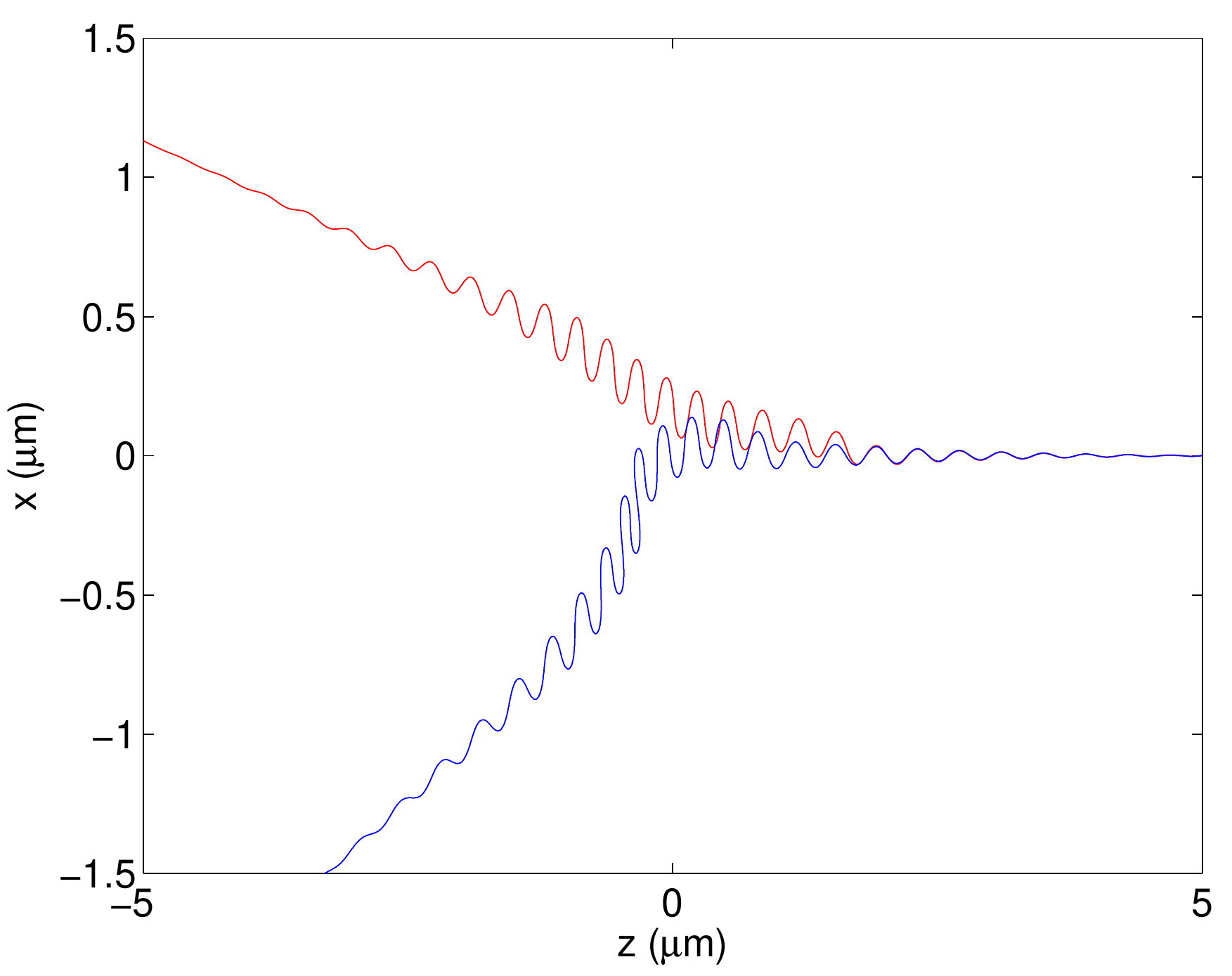}\\[2ex]
\includegraphics[width=0.7\columnwidth,clip=true,viewport=40 190 545 590]{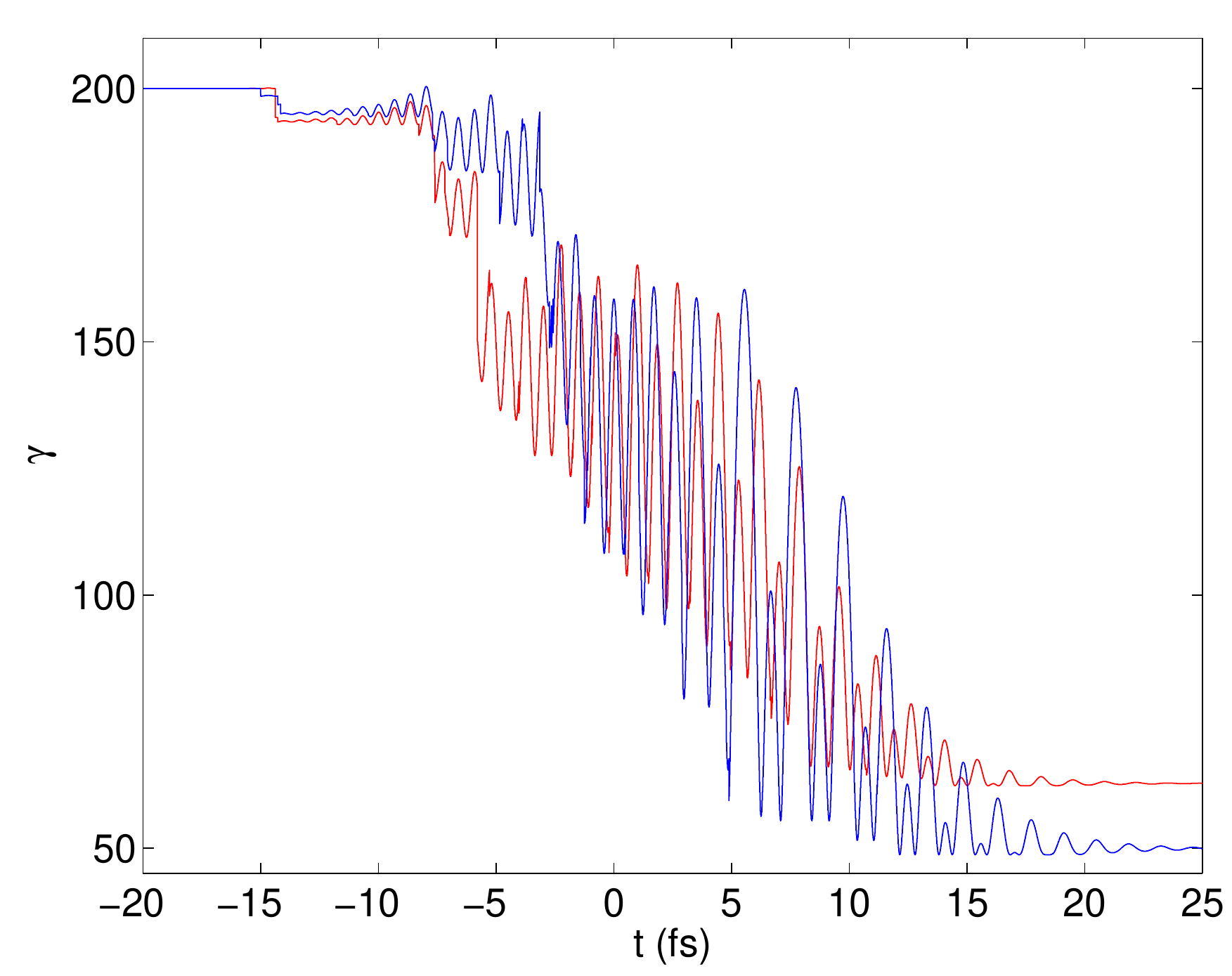}
\caption{The trajectories (top panel) and $\gamma$-factors (bottom panel) for a sample of two (QED) electrons of $\gamma_0$=200 in collision with a 30\,fs linearly polarised wave of intensity $a_0=150$ and wavelength $\lambda=0.8\,\mu$m.\label{fig:qed_traj} }
\end{figure}
\begin{figure}[htp!]
\centering
\includegraphics[width=0.7\columnwidth,clip=true,viewport=40 190 545 590]{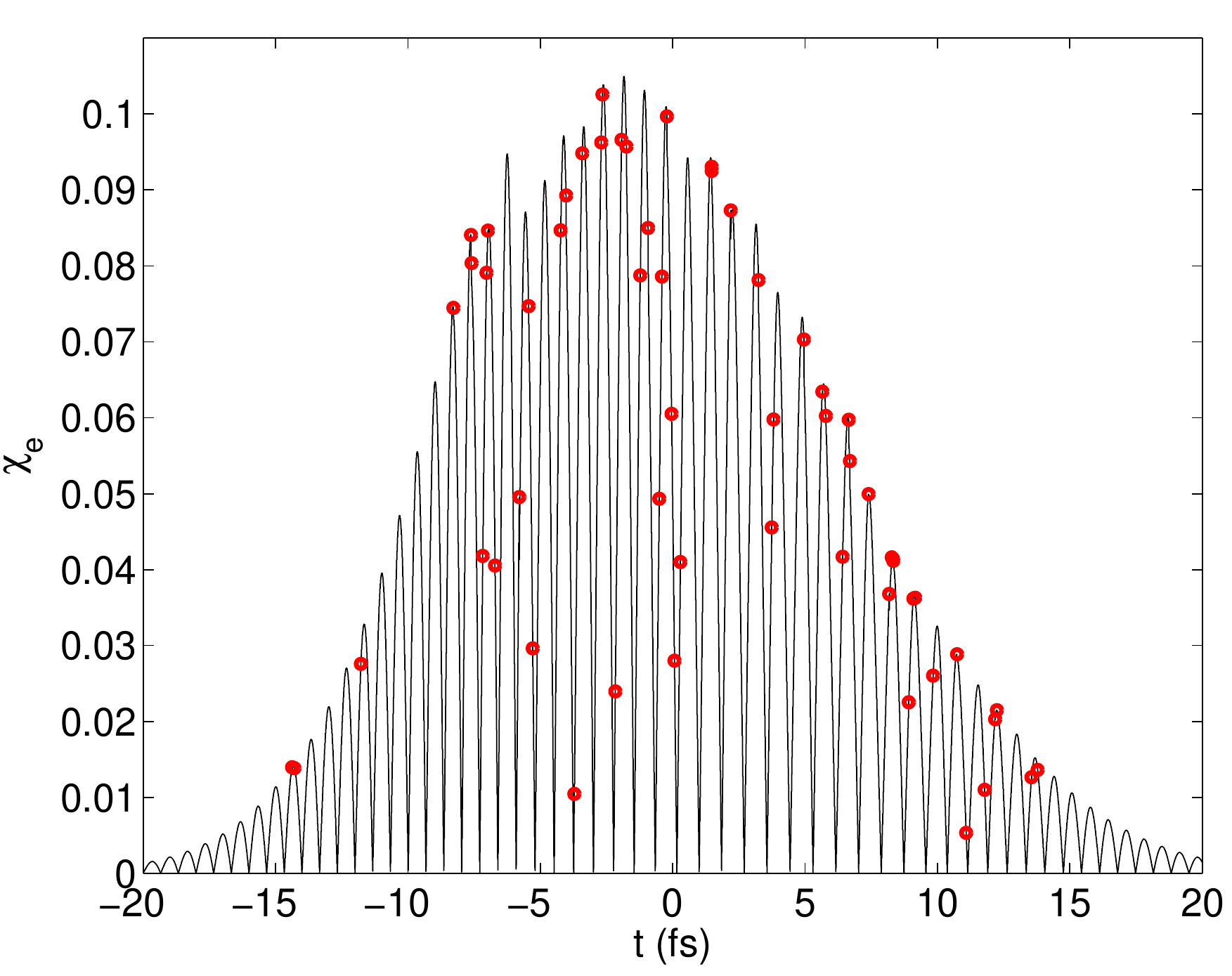}
\caption{The quantum efficiency parameter $\chi_e$ as a function of time for a single (QED) electron of $\gamma_0$=200 in collision with a 30\,fs linearly polarised wave of intensity $a_0=150$ and wavelength $\lambda=0.8\,\mu$m.  The red circles indicate when a photon emission event has taken place.\label{fig:chi_e} }
\end{figure}

In this second example let us invoke the QED routines.  Our aim is to calculate the statistical photon spectrum generated by an electron passing through an intense laser pulse.  In order to generate enough statistics to produce the spectrum we will need to execute the code a number times, always with the same initial conditions.  The easiest way to do this is to continually read in the first line of particle data in the file `particle\_input.csv'.  This is achieved by setting {\tt repeatfirstline=on}.  The file should therefore take the form
\begin{alltt}
! ...comments 
repeatfirstline=on
100
1,e,0,180,100e-6,0,0,0,200,qed,t
\end{alltt}
where we have told the code to execute 100 times using the first line of parameters as input each time.  The $\gamma$-factor has been set to 200 and we have invoked the QED routines with the flag {\tt qed}.  
The {`input\_fields.txt'} file can be similar to before, although we choose to increase the field intensity significantly, setting {\tt a0\_1=150}.  Finally, in the `input\_setup.txt' file we set {\tt QEDrecoil=on} so that every time a photon is emitted its momentum is subtracted from the electron's momentum.

Once the simulation has completed\footnote{Completion will take significantly longer than in the previous example because the electron-laser collision simulation is being executed multiple times.} the photon data will be written to the file {`photon.dat'}.  Executing the Matlab script {`photon.m'} loads various items of data into the Matlab memory and plots the photon distributions.  The photon distributions as a function of energy and emission angle are shown in Fig.~\ref{fig:photon_spectra}.  
(Note also that the final data of each particle as it leaves the {\it simulation box} has been written to the file {`final\_data.dat'}.  The Matlab script {`finaldata.m'} can be used to plot this data.)
The trajectories and $\gamma$-factors for two electrons, plotted using the script {`simlasingle.m'}, are shown in Fig.~\ref{fig:qed_traj}. The trajectory plot shows the effect of the gain in momentum transverse to the beam axis that remains post collision, as discussed in \cite{PhysRevLett.112.164801}. The plot of the $\gamma$-factor shows clearly the discontinuous changes in energy of the electron as it emits discrete photons according to Eq.~(\ref{eqn:dGam}). Finally, in Fig.~\ref{fig:chi_e} we show the parameter $\chi_e$ for a single electron. The figure shows the reduction in $\chi_e$ with each photon emission (note that the conservation law $\chi_e^{(f)}+\chi_{\gamma}=\chi_e^{(i)}$ holds, where $\chi_e^{(i)}$ and $\chi_e^{(f)}$ are the value of $\chi_e$ before and after the photon emission, respectively \cite{Ritus:1985}.) 

Note that more advanced features are discussed in the manual.

\section{Summary}\label{sec:conclusion}
We have introduced and described the software package SIMLA, designed for the study of charged-particle dynamics in laser and other background fields.  Using two simple examples to illustrate its operation, we have demonstrated how to obtain trajectories and emission spectra in classical and strong-field quantum electrodynamics, with the latter implemented via statistical routines.  More advanced features and settings are detailed in the accompanying manual.

Specifically, the code is aimed at users interested in studying situations where mutual interactions between particles can be ignored.  This includes a range of experimental set ups where high-energy particle beams are brought into collision with laser pulses, and theoretical studies, whose aim may be to understand the dynamics of single particles in different field configurations, or to test and refine the numerical methods used in larger scale simulations.

We hope that this code will prove accessible and useful to a wide range of researchers, including those who lack the specialist knowledge and super-computing infrastructure required to run large-scale `particle-in-cell' codes.

\section*{Acknowledgements}
The authors were supported by the EPSRC, grant EP/I029206/1--YOTTA.

\appendix
\section{Fields and Setup file options}
This appendix contains a summary of the input file options. There are two input files. In the first the field variables are specified, while in the second the simulation parameters are specified. 
\begin{table*}[hp!]
\caption{The variables in the input file {`input\_fields.txt'}. \label{tab:inputfields}}
\begin{tabular}{ l  l  l }
\hline
{\tt no\_fields} & INTEGER & number of background fields\\
{\tt field1,\ldots,field9} & CHAR & the type of each of the fields \\
{\tt profile1,\ldots,profile9} & CHAR & the profile of each of the fields \\
{\tt anglexz1,\ldots,anglaxz9} & REAL(8) & angles (in deg) of the field rotations in the xz-plane \\
{\tt angleyz1,\ldots,anglayz9} & REAL(8) & angles (in deg) of the field rotations in the yz-plane \\
{\tt anglexy1,\ldots,anglaxy9} & REAL(8) & angles (in deg) of the field rotations in the xy-plane \\
{\tt lambda1,\ldots,lambda9} & REAL(8) & wavelengths (in $m$) of each of the fields \\
{\tt waist1,\ldots,waist9 } & REAL(8) &  waist radii (in $m$) of each of the fields\\
{\tt a0\_1,\ldots,a0\_9} & REAL(8) & dimensionless intensity of each of the fields \\
{\tt fieldstrength1,\ldots,fieldstrength9 } & REAL(8) & strength of fields (in $eV^2$) for any constant fields\\
{\tt duration1,\ldots,duration9} & REAL(8) & durations of each of the time-dependent fields (in $s$) \\
\hline  
\end{tabular}
\vspace*{3ex}
\caption{The variables in the input file `input\_setup.txt'. \label{tab:inputsetup}}
\begin{tabular}{ l  l  l }                   
\hline
{\tt tmax} & REAL(8) & maximum time (in s) of simulation\\
{\tt xmin} &  REAL(8) &  minimum value of the $x$-coordinate (in m)\\
{\tt xmax} & REAL(8) &  maximum value of the $x$-coordinate (in m)\\
{\tt ymin} &  REAL(8) &  minimum value of the $y$-coordinate (in m)\\
{\tt ymax} & REAL(8) &  maximum value of the $y$-coordinate (in m)\\
{\tt zmin} & REAL(8) &  minimum value of the $z$-coordinate (in m)\\
{\tt zmax} &  REAL(8) &  maximum value of the $z$-coordinate (in m)\\
\hline
{\tt tminw} & REAL(8) & maximum time (in s) of simulation write-box\\
{\tt tmaxw} & REAL(8) & maximum time (in s) of simulation write-box\\
{\tt xminw} &  REAL(8) &  minimum value of the write-box $x$-coordinate (in m)\\
{\tt xmaxw} & REAL(8) &  maximum value of the write-box $x$-coordinate (in m)\\
{\tt yminw} &  REAL(8) &  minimum value of the write-box $y$-coordinate (in m)\\
{\tt ymaxw} & REAL(8) &  maximum value of the write-box $y$-coordinate (in m)\\
{\tt zminw} & REAL(8) &  minimum value of the write-box $z$-coordinate (in m)\\
{\tt zmaxw} &  REAL(8) &  maximum value of the write-box $z$-coordinate (in m)\\
\hline
{\tt solver} & CHAR &  type of numerical method\\
{\tt maxdt} & REAL(8) & maximum time step for the numerical solver (s) \\
{\tt mindt} &  REAL(8) &  minimum time step for the numerical solver (s)\\
{\tt initialdt} & REAL(8) & initial time step for the numerical solver (s) \\
{\tt grid\_err\_tol} &  REAL(8) & maximum permissible error at each time step\\
{\tt writeevery} & INTEGER & number of time steps between writing data to file \\
{\tt fileformat} & CHAR & formatting of the trajectory output files \\
{\tt QEDrecoil} & CHAR  & whether the particle recoils when emitting a photon \\
{\tt outputintensity} & CHAR &  whether to output the field intensity\\
{\tt outputfieldsdt} & REAL(8) &  timestep ($s$) between subsequent outputs of field intensity\\
{\tt fieldpointsx} & INTEGER & number of grid points for intensity output ($x$-direction) \\
{\tt fieldpointsz} & INTEGER & number of grid points for intensity output ($z$-direction) \\
\hline  
\end{tabular}
\end{table*}

\newpage
\bibliographystyle{apsrev4-1} 

%

\end{document}